\documentclass[%
aip,
pop,%
amsmath,amssymb,
reprint,%
]{revtex4-1}

\usepackage{graphicx}
\usepackage{epspdfconversion}
\usepackage{dcolumn}
\usepackage{bm}
\begin{document}

\preprint{AIP/123-QED}

\title{Anomalous diffusion due to the non-Markovian process of the dust particle velocity in complex plasmas}

\author{Z. Ghannad}
\email{z.ghannad@alzahra.ac.ir\\$\copyright$2017. This manuscript version is made available under the CC-BY-NC-ND 4.0 license \\
https://creativecommons.org/licenses/by-nc-nd/4.0/}
\author{H. Hakimi Pajouh}%
\affiliation{%
Faculty of Physics and Chemistry, Alzahra University, P. O. Box 19938-93973, Tehran, Iran
}%


\begin{abstract}
In this work, the motion of a dust particle under the influence of the random force due to dust charge fluctuations is considered as a non-Markovian stochastic process. Memory effects in the velocity process of the dust particle are studied. A model is developed based on the fractional Langevin equation for the motion of the dust grain. The fluctuation-dissipation theorem for the dust grain is derived from this equation. The mean-square displacement and the velocity autocorrelation function of the dust particle are obtained in terms of the Mittag-Leffler functions. Their asymptotic behavior and the dust particle temperature due to charge fluctuations are studied in the long-time limit. As an interesting result, it is found that the presence of memory effects in the velocity process of the dust particle as a non-Markovian process can cause an anomalous diffusion in dusty plasmas. In this case, the velocity autocorrelation function of the dust particle has a power-law decay like $t^{-\alpha-2}$, where the exponent $\alpha$ take values $0<\alpha<1$.
\begin{description}
\item[PACS numbers]
52.27.LW, 02.50.Ey
\item[Keywords]
Stochastic process, Dust charge fluctuations, Correlation function,  Dusty plasma
\end{description}
\end{abstract}

\maketitle


\section{\label{sec:level1}Introduction}
A dusty plasma consists of a mixture of electrons, ions, neutral gas atoms, and
charged dust particles. Due to the collection
of electrons and ions from the plasma, these particles gain an electric charge~\cite{Douglass}. The dust charges are not fixed, and they can fluctuate. There are two main reasons for these fluctuations. The first reason is spatial and temporal variations in plasma parameters, and the second is the discrete nature of electrons and ions, which arrive at the dust surface at random time intervals~\cite{Yaroshenko,Fortov2005}. The latter fluctuations are considered in this paper.

The influence of charge fluctuations on dynamics of dust particles in dusty plasmas has been investigated in many studies \cite{Vaulina2000,Vaulina1999,Vaulina,Quinn2000,Ivlev2010,Schmidt2015}. For example, Vaulina $etal.$ ~\cite{Vaulina} showed that charge fluctuations can be a reason for the heating of the dust particle system. Quinn and Goree~\cite{Quinn2000} developed a Langevin equation of motion to predict the dust particle temperature due to fluctuations. Ivlev $etal.$ ~\cite{Ivlev2010} proposed a novel mechanism of dust acceleration induced by charge fluctuations based on the Fokker-Planck equation. Schmidt and Piel~\cite{Schmidt2015} studied the Brownian motion of a single dust particle by charge fluctuations in the plasma sheath through the stationary Klein-Kramers equation, which is a special form of the Fokker-Planck equation~\cite{Risken}. In all these studies, the influence of charge fluctuations has been introduced as a random force in the Langevin equation or a diffusion coefficient in the Fokker-Planck equation. The Langevin or Fokker-Planck equations are built on the assumption that the fluctuations are fast, meaning that correlation time of fluctuations is much shorter than the other
characteristic times of the system, such as the relaxation time of the particle velocity. Hence, the velocity process of the particle can be considered as a Markov process with no memory, meaning that the evolution of the velocity
from time $t$ depends only on its value at this time, and not on the values it took at prior times~\cite{Pottier2009}.

The values of the relaxation timescale for dust charge fluctuations and the dust velocity entirely depend on the dusty plasma parameters. For instance, Hoang and Lazarian~\cite{Hoang} showed that the relaxation time of dust charge fluctuations $\tau_{cor}$ is comparable to the relaxation time of the dust velocity $\tau_{rel}$, i.e., $\tau_{cor}\approx\tau_{rel}$ for very small dust particles with radius $a\leq 5\times10^{-8}$ cm in the interstellar medium.
In this paper, we compare the characteristic timescales associated with the dust particle in laboratory dusty plasmas and show
depending on the plasma parameters, such as pressure or density of the neutral gas, $\tau_{cor}$ is comparable to $\tau_{rel}$. Therefore, in such situations, the main assumptions of fast charge fluctuations and the Markov process for the velocity process of the dust particle become inappropriate. Thus, memory effects in the velocity process of dust particle as a non-Markov process become important, and they cannot generally be neglected. Then, for this situation, we develop a detailed analytical model for the evolution of the dust particle exposed to the random force due to charge fluctuations based on a fractional Langevin equation. In the end, we find that the memory effects in the velocity process of the dust particle can cause an anomalous diffusion of the dust particle in dusty plasma.

This paper is organized as follows. In section \ref{Timescales}, we compare major timescales characterizing
dynamics of the dust particle and developed a model based on the fractional Langevin equation including the memory effects in the velocity of the dust particle. Section \ref{fluctuation} is devoted to derive a suitable fluctuation- dissipation theorem.
 In section \ref{calculation}, we calculate the mean-square displacement and the velocity autocorrelation function of the dust grain. In section \ref{limit} we investigate the asymptotic behavior of the results. Section \ref{summary}
 contains summary and conclusions.

\section{\label{Timescales}Theoretical model description}
In this section, we consider an isolated spherical dust particle in the sheath and study the relevant characteristic times. The particle is charged by collecting electrons and ions from the plasma. As we mentioned in the introduction, the particle charge fluctuates about the equilibrium value due to the discrete nature of the electron and ion currents.
It was shown that the autocorrelation function of dust charge fluctuations is obtained by~\cite{Khrapak1999}
\begin{equation}
\langle \delta Z(t)\delta Z(t')\rangle=\langle \delta Z^2\rangle\, \rm{exp}\left(-\beta\vert \it t-t'\vert\right),\label{eq1}
\end{equation}
where $\langle \delta Z^2\rangle$ is the square of the amplitude of random charge fluctuations, $\delta Z(t)$=$Z(t)$--$Z_0$, $Z(t)$ is the instantaneous charge, $Z_0$ is the equilibrium charge, and $\beta$ is charging frequency, which is defined as the relaxation frequency for small deviations of the charge from the equilibrium value $Z_0$. For the isolated dust particle under the condition $a\ll\lambda_D\ll l_{\rm{i(e)}}$, where $a$ is the dust radius, $\lambda_D$ is the screening length due to electrons and ions, and $ l_{\rm{i(e)}}$ is the mean free path for ion (electron) collisions with neutrals, the charging frequency can be calculated as follows~\cite{Fortov2010}
\begin{equation}
\beta=-\frac{\mathrm{d}(I_\mathrm{i}-I_\mathrm{e})}{\mathrm{d}Z}\bigg|_{Z=Z_0}=\frac{1+z}{\sqrt{2\pi}}\frac{a}{\lambda_{D\mathrm{i}}}\omega_{\mathrm{pi}}\label{eq2}
\end{equation}
where $I_\mathrm{i}$=$\sqrt{8\pi}a^2n_\mathrm{i}v_{T\mathrm{i}}\left(1-Ze^2/aT_\mathrm{i}\right)$ is the ion flux, and $I_\mathrm{e}$=$\sqrt{8\pi}a^2n_\mathrm{e}v_{T\mathrm{e}} \rm{exp}$$\left(Ze^2/aT_\mathrm{e}\right)$ is the electron flux to the particle surface obtained from the orbital- motion-limited (OML) theory, $v_{T\mathrm{i(e)}}$=$\left(T_{\mathrm{i(e)}}/m_{\mathrm{i(e)}}\right)^{1/2}$ is the ion (electron) thermal velocity, $T_{\mathrm{i(e)}}$, $m_{\mathrm{i(e)}}$, and $n_{\mathrm{i(e)}}$ are ion (electron) temperature, mass, and number density, respectively, $z$=
$\vert Z \vert e^2/aT_\mathrm{e}$ is the absolute magnitude of the particle charge in the units $aT_\mathrm{e}/e^2$, $\lambda_{D\mathrm{i}}$=$\sqrt{T_\mathrm{i}/4\pi e^2n_\mathrm{i}}$ is the ionic Debye radius, and $\omega_{\mathrm{pi}}$=$v_{T\mathrm{i}}/\lambda_{D\mathrm{i}}$ is the ion plasma frequency.

Let us now consider an isolated dust particle with fluctuating charge in the sheath. To separate the dust particle transport due to dust charge fluctuations from other processes, which in turn can influence the dust motion in the plasma sheath, we only consider the gravitational, electric field, and neutral drag forces on the dust. We hence neglect the ion drag, electron drag, and thermophoretic forces, collisions between dust particles, and other processes, which can include in more realistic models~\cite{Khrapak2002}. The motion of this dust particle is treated as a stochastic process because of the stochastic nature of the force due to dust charge fluctuations, and it can generally be modeled by a normal Langevin equation of the form~\cite{Quinn2000} 
\begin{equation}
\ddot{x}(t)+\gamma\dot{x}(t)+\omega_0^2x(t)=\xi(t),\label{eq3}
\end{equation}
where $x(t)$ is the position of the dust particle, $-\gamma \dot{x}(t)$ is the neutral drag force per unit mass,$\gamma$ is the damping
rate due to neutral gas friction, and $\omega_0$ is the oscillation frequency of the dust particle for small oscillations around a
balancing position due to the vertical confining potential~\cite{Nitter,Quinn2000}. Note that $\xi(t)$=$f(t)+\eta(t)$ is the fluctuating force per unit mass on the dust particle, which $f(t)$ is the random force per unit mass representing the effect of dust charge fluctuations, and $\eta(t)$ is the stochastic Langevin force per unit mass due to collisions with the neutral gas molecules; However, as we mentioned, due to separate the dust particle diffusion under the influence of dust charge fluctuations from other processes, we neglect the stochastic Langevin force $\eta(t)$. Therefore, in the absence of the stochastic Langevin force, $\eta(t)$=0, $\xi(t)$ reduces to the random force due to charge fluctuations, i.e., $\xi(t)$=$f(t)$.
Here, the forces acting on a dust are the electric force due to the sheath electric field, the gravitational force, i.e., $F(t)$=$F_Z(t)$+$F_g$. The electric force is given by $F_Z(t)$=$eEZ(t)$, where $Z(t)$=$Z_0$+$\delta Z(t)$ is the instantaneous dust charge, and $E$ is the electric field. Note that, for simplicity, we have neglected the fluctuations of the electric field. The force can be written as $F(t)$=$F_0$+$f(t)$, where $F_0$=$eEZ_0$+$F_g$. Thus, the random force due to dust charge fluctuations has the following form
\begin{equation}
f(t)=eE\delta Z(t).\label{eq4}
\end{equation}
In a steady state $eEZ_0$+$F_g$=$0$, so that $F(t)$=$f(t)$. By using equations (\ref{eq4}) and~(\ref{eq1}), one can see that the random force due to dust charge fluctuations per unit mass has following properties
\begin{eqnarray}
\langle f(t)\rangle=0,\nonumber
\end{eqnarray}
and
\begin{equation}
\langle f(t)f(t')\rangle=\frac{e^2E^2}{m_\mathrm{d}^2}\langle \delta Z^2\rangle\, \rm{exp}\left(-\beta\vert \it{t-t'}\vert\right),\label{eq5}
\end{equation}
where $m_\mathrm{d}$ is the dust particle mass.

Now, we study the major timescales relevant to the system under study.
In general, stochastic processes are classified into two types of Markov and non-Markov processes  based on the timescales~\cite{Ridolfi,Kim}:
\\
(1) A stochastic process is said a Markov process with no memory effects in its velocity process, if the correlation time of the random force $\tau_{cor}$ is much shorter than other characteristic times of the system, such as $\tau_s$, i.e., $\tau_{cor}\ll\tau_s$.
In this case, as we mentioned,  the evolution of the velocity at time $t$ depends only on its value at this time, and not on the values it took at prior times. \\
(2) However, if $\tau_{cor}$ is not the smallest timescale and is of the order of $\tau_s$, i.e., $\tau_{cor}\approx\tau_s$, in this regime, a stochastic process is called a non-Markov process with memory effects in its velocity process. It means that the velocity of the particle at the current time depends on its velocity at all past times.

Now, we investigate the relevant timescales in the system under study. There are three major timescales involved in dynamics of our system.
The first is the correlation time of random force due to charge fluctuations, which is obtained by equation~(\ref{eq5})
\begin{equation}
\tau_{cor}=\frac{1}{\beta}.\label{eq6}
\end{equation}
The second timescale is the oscillation time of the dust particle
\begin{equation}
\tau_{osc}=\frac{2\pi}{\omega_0},\label{eq7}
\end{equation}
which is of the order of $ 0.01-0.1$ s~\cite{Nitter,Fortov2005}. The third timescale is the relaxation time of the dust velocity
\begin{equation}
\tau_{rel}=\frac{1}{\gamma}.\label{eq8}
\end{equation}
To calculate $\tau_{cor}$ and $\tau_{rel}$, we first need to calculate the damping rate $\gamma$ and the charging frequency $\beta$.
In general, the values of the damping rate and the charging frequency entirely depend on the plasma parameters. For example, we consider two different types of plasmas with argon and krypton  neutral gases and estimate the values of $\gamma$ and $\beta$ using typical experimental values for various parameters. We assume $T_\mathrm{e}$=4 $\rm{eV}, \it{T_\mathrm{e}/T_\mathrm{i}}$=40, $n_\mathrm{i}$=$10^8 \rm{cm^{-3}}$, and the silica dust particle with radius $a$=0.5 $\mu$m and the mass density $\rho$=2 $\rm{g/cm^3}$. We also assume the neutral gases at room temperature and in the range of pressures 0.4--0.9 Torr. For argon and krypton neutral gases with $T_\mathrm{e}/T_\mathrm{i}$=40, the absolute magnitudes of the dust charge are $z\sim$2.8, and 3.2, respectively~\cite{Fortov2010,Fortov2004}. With the given parameters, we calculate the charging frequency from equation~(\ref{eq2}). The values of the correlation time of the random force due to charge fluctuations $\tau_{cor}$=$\beta^{-1}$ (in seconds) are tabulated in the last row of Table~\ref{jlab1}.

When the neutral gas mean free path is long compared to the dust grain radius, it is appropriate to use the Epstein drag force to calculate $\gamma$~\cite{Epstein,Baines}. The mean free path values of argon and krypton atoms at the maximal
pressure used (0.9 Torr) are approximately equal to 67, and 45 $\mu$m, respectively. These are about 134, and 90 times larger than the dust grain radius (0.5 $\mu$m). Thus, the damping rate is given by the Epstein formula
\begin{equation}
\gamma=\delta\sqrt{\frac{8}{\pi}}\left(\frac{T_g}{m_g}\right)^{-\frac{1}{2}}\frac{P}{\rho a},\label{eq9}
\end{equation}
where $m_g$, $T_g$, $P$ are the mass, temperature, and pressure of the neutral gas, respectively. The parameter $\delta$ is 1.39 for diffuse reflection or 1 for specular reflection of the neutral gas atoms from the dust particle~\cite{Epstein,Baines}. We use $\delta$=1.39 and the given parameters to calculate the damping rate from equation~(\ref{eq9}). The relaxation time values of the dust velocity 
$\tau_{rel}$=$\gamma^{-1}$ (in seconds) for each gas at various pressures are listed in Table~\ref{jlab1}. As shown in Table~\ref{jlab1}, 
upon increasing the pressure (or equivalently increasing the density) of the neutral gas, the relaxation time of the dust velocity becomes comparable to the relaxation time of the random force. Thus, in this situation, $\tau_{cor}$ is not the smallest timescale of the system under study, and is of the order of $\tau_{rel}$, i.e., $\tau_{cor}\approx\tau_{rel}$, and memory effects in the velocity of the dust particle cannot be neglected. As a result, in this regime, the normal Langevin equation becomes inappropriate for the description of the dust motion because this equation is built on the Markovian assumption with no memory. When  $\tau_{rel}$ becomes comparable to $\tau_{cor}$, the memory effects in the relaxation of the dust velocity become important, meaning that the velocity of the dust particle at the current time depends on its velocity at past times, and it means that the process of the dust velocity relaxation is retarded, and these retarded effects (or equivalently memory effects) are characterized with memory kernel in the friction force within Langevin equation. Now, this equation with retarded friction is called the fractional Langevin equation. Hence, the fractional Langevin equation is built on the non-Markovian assumption, while the Langevin equation without the existence a retarded friction is built on the Markov assumption. 
\begin{table}
\caption{\label{jlab1}The values of the relaxation time of the dust velocity $\tau_{rel}$ for each gas at various pressures.}
\begin{center}
\footnotesize
\bgroup
\def\arraystretch{1.7}
\begin{tabular}{ccccccc}
\hline
Gas &Kr&Ar\\
\hline
\,\,\,\,$P$=\,0.4 Torr&\,\,\,\,\,\,\,\,\,\,1.45$\times$$10^{-3}\,\, \mathrm{s}$&\,\,\,\,\,\,\,\,\,\,2.10$\times$$10^{-3}  \,\,\mathrm{s}$\\
\,\,\,\,$P$=\,0.7 Torr&\,\,\,\,\,\,\,\,\,\,8.31$\times$$10^{-4}\,\, \mathrm{s}$&\,\,\,\,\,\,\,\,\,\,1.20$\times$$10^{-3} \,\, \mathrm{s}$ \\
\,\,\,\,$P$=\,0.9 Torr&\,\,\,\,\,\,\,\,\,\,6.46$\times$$10^{-4} \,\,\mathrm{s}$ &\,\,\,\,\,\,\,\,\,\,9.35$\times$$10^{-4} \,\, \mathrm{s}$\\
\hline
$\tau_{cor}$&\,\,\,\,\,\,\,\,\,1.86$\times$$10^{-4 \,\,} \mathrm{s}$&\,\,\,\,\,\,\,\,\,\,1.42$\times$$10^{-4}\,\,\mathrm{s}$\\
\hline
\end{tabular}
\egroup
\end{center}
\end{table}\\
As shown in Table~\ref{jlab1}, memory effects can be clearly seen at neutral gas pressures about 1 Torr and higher, $P\geq$1 Torr. It is worth mentioning that in obtaining the expression for charging frequency from equation~(\ref{eq2}), we have neglected ion-neutral collisions and have obtained the charging currents from the OML theory. However, ion-neutral collisions (neglected in the OML theory) significantly affect the charging process even in the regime of weak ion collisionality ($l_\mathrm{i}>\lambda_D$), relevant to the majority
of complex plasmas experiments in gas discharges~\cite{Khrapak2012,Khrapak2009}. Therefore, ion-neutral collisions should be included in more realistic models, which in turn can affect the magnitude of charging frequency.

Now,  we study the motion of the dust grain in the regime $\tau_{cor}\approx\tau_{rel}$ and consider memory effects in the velocity of an isolated dust  particle exposed to the random force due to dust charge fluctuations. We develop a model  based on the fractional Langevin equation (FLE) for the motion of the dust grain because this equation includes the memory kernel function, which is non-local in time and shows memory effects in the velocity of the dust particle. The FLE is in the following form~\cite{Burov,Ghannad}
\begin{eqnarray}
&\ddot{x}(t)+\frac{\bar{\gamma}}{\Gamma(1-{\alpha})}\int_0^t \left(\frac{\vert t-t'\vert}{\tau_{cor}}\right)^{-\alpha}\dot{x}(t')\mathrm{d}t'+\omega_0^2x(t)\nonumber\\
&=f(t)\nonumber
\end{eqnarray}
where 0$<\alpha<$1, and $\Gamma(1-{\alpha})$ is the gamma function. $\bar{\gamma}$ is the scaling factor with physical dimension (time)$^{-2}$ and must be introduced to ensure the correct dimension of the equation. We define $\bar{\gamma}=(\tau_{rel}\tau_{cor})^{-1}$, so that for $\alpha$=1, and according to the Dirac generalized function, $\delta(t-t')$=$\vert t-t'\vert^{-1}/\Gamma(0)$~\cite{Gel'fand}, the FLE equation reduces to the Langevin equation~(\ref{eq3}).
\\
The FLE equation can be rewritten in the following form
\begin{equation}
\ddot{x}(t)+\int_0^t\gamma(t-t')\dot{x}(t')\mathrm{d}t'+\omega_0^2x(t)=f(t),\label{eq10}
\end{equation}
where $\gamma(t-t')$ is called  the memory kernel function
\begin{equation}
\gamma(t-t')=\frac{(\tau_{rel}\tau_{cor})^{-1}}{\Gamma(1-\alpha)}\left(\frac{\vert t-t'\vert}{\tau_{cor}}\right)^{-\alpha}.\label{eq11}
\end{equation}
As interesting feature, the name fractional in the fractional Langevin equation is due to the fractional derivative, which is defined by~\cite{Caputo}
\begin{eqnarray}
\frac{\mathrm{d}^{\alpha}f(t)}{\mathrm{d}t^{\alpha}}={_0}{\mathcal{D}}{_t^{\alpha-1}}\left(\frac{\mathrm{d}f(t)}{\mathrm{d}t}\right),\nonumber
\end{eqnarray}
where ${_0}{\mathcal{D}}{_t^{\alpha-1}}$ is the Riemann-Liouville fractional operator~\cite{Miller,Podlubny}
\begin{eqnarray}
{_0}{\mathcal{D}}{_t^{\alpha-1}}f(t)=\frac{1}{\Gamma(1-\alpha)}\int_0^t\left(t-t'\right)^{-\alpha}f(t)\mathrm{d}t.\nonumber
\end{eqnarray}
As a result, the fractional Langevin equation reads
\begin{eqnarray}
\ddot{x}(t)+\frac{\bar{\gamma}}{\tau_{cor}^{-\alpha}}\frac{\mathrm{d}^{\alpha}x(t)}{\mathrm{d}t^{\alpha}}+\omega_0^2x(t)=f(t);\nonumber
\end{eqnarray}
therefore, the name fractional Langevin equation is confirmed. As we see, for $\alpha$=1, this equation reduces to the normal Langevin equation ~(\ref{eq3}).

As shown, considering the memory effects in the velocity process of the particle leads to a modification of the 
Langevin equation. In addition, this consideration leads to a suitable relation between the memory kernel and the autocorrelation function of the random force; Hence, in section \ref{fluctuation}, we derive this relation.
\section{\label{fluctuation}The fluctuation-dissipation theorem}
To obtain a suitable relation between the memory kernel and the autocorrelation function of the random force, we first define the Fourier transforms for the position and random force as follows
\begin{equation}
x(\omega)=\int_{-\infty}^{\infty}x(t)\,e^{\mathrm{i}\omega t}\mathrm{d}t,\,\,\,\,f(\omega)=\int_{-\infty}^{\infty}f(t)\,e^{\mathrm{i}\omega t}\mathrm{d}t.\label{eq12}
\end{equation}
Using equation~(\ref{eq12}) and taking the Fourier transform of equation~(\ref{eq10}) yields
\begin{equation}
x(\omega)=\frac{f(\omega)}{\omega_0^2-\omega^2-\mathrm{i}\omega\gamma(\omega)},\label{eq13}
\end{equation}
where $\gamma(\omega)$, defined by $\gamma(\omega)=\int_0^\infty \gamma(t)\,e^{\mathrm{i}\omega t}\mathrm{d}t$,
denotes the Fourier-Laplace transform of the memory kernel $\gamma(t)$.
Then, by using $v(\omega)=-\mathrm{i}\omega x(\omega)$, we have
\begin{equation}
v(\omega)=\frac{-\mathrm{i}\omega f(\omega)}{\omega_0^2-\omega^2-\mathrm{i}\omega\gamma(\omega)}.\label{eq14}
\end{equation}
The velocity autocorrelation function (VACF) is defined by
\begin{equation}
C_\mathrm{v}(\tau)=\langle v(t)v(t+\tau)\rangle=\lim_{\theta \rightarrow \infty}\frac{1}{\theta}\int_{-\theta/2}^{\theta/2}v(t)v(t+\tau)\mathrm{d}t\label{eq15}
\end{equation}
where $\theta$ is the time interval for the integration. The power spectrum $S(\omega)$ and the autocorrelation function of a stochastic process $C(\tau)$ are related by the Wiener-Khintchine theorem~\cite{Yates}
\begin{equation}
S(\omega)=\int_{-\infty}^{\infty}C(\tau)\,\mathrm{e}^{\mathrm{i}\omega \tau}\mathrm{d}\tau,\label{eq16}
\end{equation}
\begin{equation}
C(\tau)=\frac{1}{2\pi}\int_{-\infty}^{\infty}S(\omega)\,\mathrm{e}^{-\mathrm{i}\omega \tau}\mathrm{d}\omega.\label{eq17}
\end{equation}
By using equations~(\ref{eq15}), (\ref{eq16}), and the Fourier transform of the velocity, $v(\omega)$=$\int_{-\infty}^{\infty}v(t)\,\mathrm{e}^{\mathrm{i}\omega t}\mathrm{d}t$, we obtain the velocity power spectrum (i.e., the Fourier transform of the velocity autocorrelation function) as
\begin{equation}
S_\mathrm{v}(\omega)=\lim_{\theta \rightarrow \infty}\frac{1}{\theta}\vert v(\omega)\vert^2;\label{eq18}
\end{equation}
then, by using equation~(\ref{eq14}), we obtain the relation between the power spectrums of the velocity and random force
\begin{equation}
S_\mathrm{v}(\omega)=\frac{\omega^2 S_{\mathrm{f}}(\omega)}{\vert \omega_0^2-\omega^2-\mathrm{i}\omega\gamma(\omega)\vert^2},\label{eq19}
\end{equation}
where $S_{\mathrm{f}}(\omega)$=$\lim_{\theta \rightarrow \infty}\frac{1}{\theta}\vert f(\omega)\vert^2$ is the power spectrum of the random force. Note that equation~(\ref{eq19}) was obtained by using the fractional Langevin equation for the non-Markov velocity process.
Similarly, we obtain the relation between the power spectrums of the random force and velocity for the Markov velocity process by using the Langevin equation~(\ref{eq3})  in the following form
\begin{equation}
S_v(\omega)=\frac{\omega^2 S_{\mathrm{f}}(\omega)}{\vert \omega_0^2-\omega^2-\mathrm{i}\omega\gamma\vert^2}.\label{eq20}
\end{equation}
When the power spectrum of the random force $S_{\mathrm{f}}(\omega)$ is given, the above equation yields the velocity power spectrum $S_v(\omega)$ from which the VACF is obtained by equation~(\ref{eq17}). If VACF should include the velocity in the thermal equilibrium, $S_{\mathrm{f}}(\omega)$ is required to satisfy a special condition. Now, we obtain this condition. In the long-time limit $(\vert t-t'\vert\gg\tau_{cor})$, the autocorrelation function of the random force obtained from equation~(\ref{eq5}) reduces to 
\begin{equation}
C_f(\tau)=\langle f(t)f(t+\tau)\rangle=\frac{e^2E^2}{m_{\mathrm{d}}^2\beta}\langle \delta Z^2\rangle \delta(\tau);\label{eq21}
\end{equation}
then, by using equation~(\ref{eq16}), the force power spectrum reads
\begin{equation}
S_{\mathrm{f}}=\frac{e^2E^2}{m_{\mathrm{d}}^2\beta}\langle \delta Z^2\rangle.\label{eq22}
\end{equation}
The velocity autocorrelation function can be obtained by using equations~(\ref{eq22}),~(\ref{eq20}), and~(\ref{eq17})
\begin{eqnarray}
C_{\mathrm{v}}(\tau)=\frac{S_{\mathrm{f}}}{2\pi}\int_{-\infty}^{\infty}\frac{\omega^2\mathrm{e}^{-i\omega\tau}}{(\omega_0^2-\omega^2)^2+\gamma^2\omega^2}\mathrm{d}\omega\nonumber
\end{eqnarray}
\begin{equation}
=\frac{S_{\mathrm{f}}}{2\gamma}\mathrm{e}^{-\gamma\tau/2}\left(\rm{cos}(\omega_1\tau)-\frac{\gamma}{2\omega_1}\rm{sin}(\omega_1\tau)\right),\label{eq23}
\end{equation}
where $\omega_1^2=\omega_0^2-\gamma^2/4$.
The mean-square velocity $\langle v^2\rangle$ can be evaluated by substituting $\tau$=0 into equation~(\ref{eq23})
\begin{equation}
\langle v^2\rangle=\frac{S_{\mathrm{f}}}{2\gamma}=\frac{e^2E^2\langle\delta Z^2\rangle}{2m_{\mathrm{d}}^2\beta\gamma}.\label{eq24}
\end{equation}
The mean-square velocity in the long-time limit (the equilibrium velocity) is representative of the dust temperature
\begin{equation}
T_{\mathrm{d}}=m_{\mathrm{d}}\langle v^2\rangle.\label{eq25}
\end{equation}
Thus, by using equations (\ref{eq24}) and (\ref{eq25}), the special condition for the power spectrum of the random force $S_{\mathrm{f}}(\omega)$ is obtained by
\begin{equation}
S_{\mathrm{f}}=\frac{2T_{\mathrm{d}}\gamma}{m_{\mathrm{d}}}.\label{eq26}
\end{equation}
Note that the above condition (equation~ (\ref{eq26})) was obtained by using equation~(\ref{eq20}) for the Markov velocity process. For the non-Markov velocity process given by the FLE , the relation between the power spectrums of the velocity and random force is given by equation~(\ref{eq19}) instead of~(\ref{eq20}); hence, the condition for the power spectrum $S_{\mathrm{f}}(\omega)$, in this case, is a generalization of equation~(\ref{eq26}) as follows
\begin{equation}
S_{\mathrm{f}}(\omega)=\frac{2T_{\mathrm{d}}}{m_{\mathrm{d}}}\rm{Re}\left(\gamma(\omega)\right);\label{eq27}
\end{equation}
then, by using equations (\ref{eq12}), (\ref{eq16}), and (\ref{eq27}), we obtain
\begin{eqnarray}
\langle f(\omega)f^*(\omega')\rangle=2\pi S_{\mathrm{f}}(\omega)\delta(\omega-\omega')\nonumber
\end{eqnarray}
\begin{equation}
\,\,\,\,\,\,\,\,\,\,\,\,\,\,\,\,\,\,\,\,\,\,\,\,\,\,\,\,\,\,\,\,\,\,\,\,=\frac{4\pi T_{\mathrm{d}}}{m_{\mathrm{d}}}\rm{Re}\left(\gamma(\omega)\delta(\omega-\omega')\right),\label{eq28}
\end{equation}
where $f^*(\omega')$ is the complex conjugate of the function $f(\omega')$. Therefore, we obtain the relation between the autocorrelation function of the random force and the memory kernel by using the inverse Fourier transform of  equation~(\ref{eq28})
\begin{equation}
\langle f(t)f(t')\rangle=\frac{T_{\mathrm{d}}}{m_{\mathrm{d}}}\gamma(t-t')=\frac{S_{\mathrm{f}}}{2\gamma}\gamma(t-t').\label{eq29}
\end{equation}
We call this the fluctuation-dissipation theorem for the dust particle because dust charge fluctuations ($S_{\mathrm{f}}\propto\langle \delta Z^2\rangle$, according to equation~(\ref{eq22})) are necessarily accompanied by the friction ($\gamma$ in the denominator).

Now, we solve equation~(\ref{eq10}) with the initial conditions $x_0$=$x(0)$ and $v_0$=$v(0)$, by using the Laplace transform technique~\cite{Grebenkov,Vinales2006}. Taking the Laplace transform of equation~(\ref{eq10}) yields
\begin{equation}
\hat{x}(s)=x_0\frac{1-\omega_0^2\hat{G}(s)}{s}+\left(v_0+\hat{f}(s)\right)\hat{G}(s),\label{eq30}
\end{equation}
where
\begin{equation}
\hat{G}(s)=\frac{1}{s^2+s\hat{\gamma}(s)+\omega_0^2},\label{eq31}
\end{equation}
where $\hat{\gamma}(s)$=$\gamma(\tau_{cor}s)^{\alpha-1}$ is the Laplace transform of the memory kernel $\gamma(t)$, and the Laplace transform of the function $f(t)$ is defined by $\hat{f}(s)$=$\int_0^\infty f(t)\mathrm{e}^{-\it{st}}\mathrm{d}t$. The inverse Laplace transform of equation~(\ref{eq30}) yields
\begin{equation}
x(t)=\langle x(t)\rangle+\int_0^t G(t-t')f(t')\mathrm{d}t',\label{eq32}
\end{equation}
where $\langle x(t)\rangle$=$v_0G(t)+x_0\left(1-\omega_0^2I(t)\right)$ is the mean dust position, and the function $G(t)$ is the inverse Laplace transform of $\hat{G}(s)$. The function $I(t)$ is given by
\begin{equation}
I(t)=\int_0^t G(t')\mathrm{d}t'.\label{eq33}
\end{equation}

The dust particle velocity is the derivative of the dust position; therefore, by using equations~(\ref{eq32}) and~(\ref{eq33}), we find
\begin{equation}
v(t)=\langle v(t)\rangle+\int_0^tg(t-t')f(t')\mathrm{d}t',\label{eq34}
\end{equation}
where $\langle v(t)\rangle$=$v_0g(t)-\omega_0^2x_0G(t)$ is the mean dust velocity, and $g(t)$=$\mathrm{d}G(t)/\mathrm{d}t$. The functions $x(t)$ and $v(t)$ help us find the mean-square displacement (MSD) and the velocity autocorrelation function of the dust particle. In section \ref{calculation}, we calculate the MSD and VACF for the dust particle.

\section{\label{calculation}Calculations of the MSD and VACF}
MSD and VACF functions are two useful quantitative tools to
investigate random processes. The MSD of a dust particle is obtained by the formula
\begin{equation}
MSD(t)=\lim_{t_0 \rightarrow \infty}\langle [x(t+t_0)-x(t_0)]^2\rangle,\label{eq35}
\end{equation}
where $x(t+t_0)$ and $x(t_0)$ are the positions of the dust particle at the two time points $t+t_0$ and $t_0$, respectively, and $\langle...\rangle$ denotes an
average over an ensemble of random trajectories. To calculate MSD, we first calculate the two-point correlation function $\langle x(t_1)x(t_2)\rangle$ by using equation~(\ref{eq32}), and we obtain
\begin{eqnarray}
\langle x(t_1)x(t_2)\rangle=\langle x(t_1)\rangle\langle x(t_2)\rangle+\int_0^{t_1}\mathrm{d}t'_1G(t_1-t'_1)\nonumber
\end{eqnarray}
\begin{equation}
\,\,\,\,\,\,\,\,\,\,\,\,\,\,\,\,\,\times\int_0^{t_2}\mathrm{d}t'_2 G(t_2-t'_2)\langle f(t'_1)f(t'_2)\rangle.\label{eq36}
\end{equation}
By using equations~(\ref{eq36}),~(\ref{eq29}),~(\ref{eq24}), and some calculations, we obtain
\begin{eqnarray}
\langle x(t_1)x(t_2)\rangle=\langle x(t_1)\rangle\langle x(t_2)\rangle+\frac{e^2E^2\langle \delta z^2\rangle}{2m_{\mathrm{d}}^2\beta\gamma}(I(t_1)+I(t_2)\nonumber
\end{eqnarray}
\begin{equation}
 -I(\vert t_1-t_2\vert)-\omega_0^2I(t_1)I(t_2)-G(t_1)G(t_2)).\label{eq37}
\end{equation}
Finally, by using equations~(\ref{eq37}) and~(\ref{eq35}), one reads
\begin{eqnarray}
MSD(t)=\lim_{t_0\to\infty}\left\lbrace\frac{e^2E^2\langle \delta z^2\rangle}{m_{\mathrm{d}}^2\beta\gamma}I(t)+\left(v_0^2-\frac{e^2E^2\langle \delta z^2\rangle}{2m_{\mathrm{d}}^2\beta\gamma}\right)\right.\nonumber
\end{eqnarray}
\begin{eqnarray}
\times(G(t_0)-G(t+t_0))^2+\left(x_0^2\omega_0^4-\omega_0^2\frac{e^2E^2\langle \delta z^2\rangle}{2m_{\mathrm{d}}^2\beta\gamma}\right)\nonumber
\end{eqnarray}
\begin{eqnarray}
\times(I(t_0)-I(t+t_0))^2-2x_0v_0\omega_0^2(I(t_0)-I(t+t_0))\nonumber
\end{eqnarray}
\begin{equation}
\left.\,\,\,\,\times(G(t_0)-G(t+t_0))\right\rbrace\label{eq38}
\end{equation}
To calculate the function $I(t)$ and $G(t)$ in limit $t \to \infty$, by using the final value theorem~\cite{Beerends,Vinales2009}, one finds
\begin{equation}
\lim_{t \rightarrow \infty}I(t)=\lim_{s \rightarrow 0}s \hat{I}(s)\label{eq39}
\end{equation}
where $\hat{I}(s)$=$\hat{G}(s)/s$ is the Laplace transform of $I(t)$ and is obtained by the Laplace transform of equation~(\ref{eq33}). Then, by using equation~(\ref{eq31}), one reads
\begin{equation}
\hat{I}(s)=\frac{\hat{G}(s)}{s}=\frac{s^{-1}}{s^2+\gamma\tau_{cor}^{\alpha-1}s^{\alpha}+\omega_0^2}.\label{eq40}
\end{equation}
Therefore, by using equations~(\ref{eq39}) and~(\ref{eq40}), one reads
\begin{equation}
\lim_{t \rightarrow \infty}I(t)=\lim_{s \rightarrow 0}\frac{1}{s^2+\gamma\tau_{cor}^{\alpha-1}s^{\alpha}+\omega_0^2}=\frac{1}{\omega_0^2}.\label{eq41}
\end{equation}
Similarly, through equation~(\ref{eq31}), one reads
\begin{equation}
\lim_{t \rightarrow \infty}G(t)=\lim_{s \rightarrow 0}s \hat{G}(s)=\lim_{s \rightarrow 0}\frac{s}{s^2+s\hat{\gamma}(s)+\omega_0^2}=0\label{eq42}
\end{equation}
By substituting equations~(\ref{eq41}) and~(\ref{eq42}) into equation~(\ref{eq38}), the MSD is obtained
\begin{equation}
MSD(t)=\frac{e^2E^2\langle \delta z^2\rangle}{m_{\mathrm{d}}^2\beta\gamma}I(t)\label{eq43}
\end{equation}
To calculate the function $I(t)$, we rewrite equation~(\ref{eq40}) in the following series~\cite{Podlubny}
\begin{equation}
\hat{I}(s)=\sum_{n=0}^\infty(-1)^n\omega_0^{2n}\frac{s^{-\alpha n-\alpha-1}}{\left(s^{2-\alpha}+\gamma\tau_{cor}^{\alpha-1}\right)^{n+1}}.\label{eq44}
\end{equation}
By applying the inverse Laplace transform of equation~(\ref{eq44}) and using the following relation~\cite{Podlubny}
\begin{eqnarray}
\int_0^\infty \mathrm{e}^{-st}t^{\alpha n+\beta-1}E_{\alpha,\beta}^{(n)}\left(\pm at^{\alpha}\right)\mathrm{d}t=\frac{n!s^{\alpha-\beta}}{\left(s^{\alpha}\mp a\right)^{n+1}},\nonumber
\end{eqnarray}
we obtain
\begin{eqnarray}
&I(t)=\sum_{n=0}^\infty\frac{(-1)^n}{n!}\omega_0^{2n}t^{2n+2}\times\nonumber\\
&E_{2-\alpha,3+\alpha n}^{(n)}\left(-\gamma\tau_{cor}(t/\tau_{cor})^{2-\alpha}\right),\label{eq45}
\end{eqnarray}
where $E_{\alpha,\beta}(z)$ is the generalized Mittag-Leffler (ML) function
\begin{eqnarray}
E_{\alpha,\beta}(z)=\sum_{n=0}^\infty\frac{z^n}{\Gamma(\alpha n+\beta)}\,\,\,\,\,\,\,\alpha,\beta>0,\nonumber
\end{eqnarray}
and for $\beta$=1, reduces to
\begin{equation}
E_{\alpha}(z)=\sum_{n=0}^\infty\frac{z^n}{\Gamma(1+\alpha n)}\,\,\,\,\,\,\,\alpha>0,\label{eq46}
\end{equation}
where is called Mittag-Leffler function.
This function is a direct generalization of the exponential function. For $\alpha$=1, we have the exponential function, $E_{1}(z)$=$\mathrm{e}^{z}$. The function $E_{\alpha,\beta}^{(n)}(z)$ is the derivative of the generalized ML function
\begin{equation}
E_{\alpha,\beta}^{(n)}(z)=\frac{d^n}{dz^n}E_{\alpha,\beta}(z).\label{eq47}
\end{equation}

The VACF of the dust grain, is defined by~\cite{Vinales2006}
\begin{equation}
C_{\mathrm{v}}(t)=\lim_{t_0 \rightarrow \infty}\langle v(t_0)v(t+t_0)\rangle.\label{eq48}
\end{equation}
To calculate VACF, we first calculate the function $\langle v(t_1)v(t_2)\rangle$ by the second-order partial derivative of the two-point correlation function $\langle x(t_1)x(t_2)\rangle$ from equation~(\ref{eq37}), and we obtain
\begin{eqnarray}
\langle v(t_1)v(t_2)\rangle=\frac{\partial^2}{\partial t_2\partial t_1}\left[\langle x(t_1)x(t_2)\rangle\right]=\langle \dot{x}(t_1)\rangle\langle \dot{x}(t_2)\rangle+\nonumber
\end{eqnarray}
\begin{equation}
 \frac{e^2E^2\langle \delta z^2\rangle}{2m_{\mathrm{d}}^2\beta\gamma}(g(\vert t_1-t_2\vert)-\omega_0^2G(t_1)G(t_2)-g(t_1)g(t_2)).\label{eq49}
\end{equation}
 By substituting $t_1=t_0$ and $t_2=t+t_0$ in equation~(\ref{eq49}); then, by substituting equation~(\ref{eq49}) in equation~(\ref{eq48}), we have
\begin{eqnarray}
C_{\mathrm{v}}(t)=\lim_{t_0\to\infty}\left\lbrace\langle \dot{x}(t_0)\rangle\langle \dot{x}(t+t_0)\rangle+ \frac{e^2E^2\langle \delta z^2\rangle}{2m_{\mathrm{d}}^2\beta\gamma}\times\right.\nonumber
\end{eqnarray}
\begin{equation}
\left.\,\,\,\,\,\,\left(g(t)-\omega^2_0G(t_0)G(t+t_0)-g(t_0)g(t+t_0)\right)\right\rbrace.\label{eq50}
\end{equation}
By using the final value theorem for $\hat{g}(s)$=$s\hat{G}(s)$, and equation~(\ref{eq31}), we get
\begin{equation}
\lim_{t \rightarrow \infty}g(t)=\lim_{s \rightarrow 0}\frac{s^2}{s^2+\gamma\tau_{cor}^{\alpha-1}s^{\alpha}+\omega_0^2}=0;\label{eq51}
\end{equation}
then, by substituting equations~(\ref{eq51}),~(\ref{eq42}) and $\langle\dot{x}(t_0)\rangle$=\\$v_0g(t_0)-\omega^2_0x_0G(t_0)$ into~(\ref{eq50}), the velocity autocorrelation function of the dust particle is obtained 
\begin{equation}
C_{\mathrm{v}}(t)=\frac{e^2E^2\langle \delta z^2\rangle}{2m_{\mathrm{d}}^2\beta\gamma}g(t).\label{eq52}
\end{equation}
VACF can also be obtained from MSD. The mean-square displacement of the particle is related to its velocity autocorrelation function using the following relation:
\begin{equation}
MSD(t)=2\int_0^t(t-t')C_{\rm{v}}(t')\mathrm{d}t'\label{eq53}
\end{equation}
or equivalently
\begin{equation}
C_{\rm{v}}(t)=\frac{1}{2}\frac{\mathrm{d}^2}{\mathrm{d}t^2}MSD(t).\label{eq54}
\end{equation}
by substituting equation~(\ref{eq43}) into this equation, VACF is obtained
\begin{eqnarray}
C_{\rm{v}}(t)=\frac{e^2E^2\langle \delta z^2\rangle}{2m_{\mathrm{d}}^2\beta\gamma}\frac{\mathrm{d}^2}{\mathrm{d}t^2}I(t)=\frac{e^2E^2\langle \delta z^2\rangle}{2m_{\mathrm{d}}^2\beta\gamma}g(t),\nonumber
\end{eqnarray}
where $g(t)=\mathrm{d}G/\mathrm{d}t=\mathrm{d^2}I/\mathrm{d}t^2$ has been used.
To calculate the function $g(t)$ and $G(t)$ in terms of the ML function, we use the following identity~\cite{Haubold}
\begin{equation}
\frac{d}{dt}t^{\beta-1}E_{\alpha,\beta}\left(at^{\alpha}\right)=t^{\beta-2}E_{\alpha,\beta-1}\left(at^{\alpha}\right)\label{eq55};
\end{equation}
then, by using equations~(\ref{eq45}), ~(\ref{eq55}), and using $g(t)$=$dG(t)/dt$, we get
\begin{eqnarray}
G(t)=&\sum_{n=0}^\infty \frac{(-1)^n}{n!}\omega_0^{2n}t^{2n+1}\times\nonumber\\
&E_{2-\alpha,2+\alpha n}^{(n)}\left(-\gamma\tau_{cor}(t/\tau_{cor})^{2-\alpha}\right),\label{eq56}
\end{eqnarray}
and
\begin{eqnarray}
g(t)=&\sum_{n=0}^\infty \frac{(-1)^n}{n!}\omega^{2n}_0t^{2n}\times\nonumber\\
&E_{2-\alpha,1+\alpha n}^{(n)}\left(-\gamma\tau_{cor}(t/\tau_{cor})^{2-\alpha}\right).\label{eq57}
\end{eqnarray}
By substituting equation~(\ref{eq57}) into equation~(\ref{eq52}), the VACF of dust particle can be written as
\begin{eqnarray}
C_{\mathrm{v}}(t)=\frac{e^2E^2\langle \delta z^2\rangle}{2m_{\mathrm{d}}^2\beta\gamma}\sum_{n=0}^\infty \frac{(-1)^n}{n!}(\omega_0t)^{2n}\times\nonumber
\end{eqnarray}
\begin{equation}
\,\,\,\,\,\,\,\,\,\,\,\,\,\,\,\,\,\,\,\,\,\,E_{2-\alpha,1+\alpha n}^{(n)}\left(-\gamma\tau_{cor}(t/\tau_{cor})^{2-\alpha}\right).\label{eq58}
\end{equation}
In section \ref{limit}, we investigate the asymptotic behaviors of the MSD and VACF of dust particle to gain a physical insight.
\section{\label{limit}Asymptotic behaviors of results}
Now, we investigate the long-time behavior of the MSD and VACF. We first study the asymptotic behavior of the function $I(t)$ for $t\gg\tau_{cor}$.
 The asymptotic behavior of the Mittag-Leffler functions is given by~\cite{Erdelyi}
\begin{equation}
E_{\alpha}(y)\sim-\frac{y^{-1}}{\Gamma(1-\alpha)},\,\,\,y\to\infty,\label{eq59}
\end{equation}
\begin{equation}
E_{\alpha,\beta}(y)\sim-\frac{y^{-1}}{\Gamma(\beta-\alpha)},\,\,\,y\to\infty.\label{eq60}
\end{equation}
By substituting equation~(\ref{eq47}) into equation~(\ref{eq45}) and using equations~(\ref{eq43}),~(\ref{eq59}) and~(\ref{eq60}), and some calculations, the mean-square displacement of the dust particle is obtained by
\begin{eqnarray}
MSD(t)\sim-\frac{e^2E^2\langle \delta z^2\rangle}{m_{\mathrm{d}}^2\beta\gamma\omega^2_0}‎\frac{1}{‎\Gamma‎(1-‎\alpha‎)}‎\frac{‎\gamma‎\tau_{cor}^{-1}}{‎\omega‎^2_0}‎
‎\left(‎‎‎\frac{t}{‎\tau‎_{cor}}‎\right)‎^{-‎\alpha‎}‎‎‎‎\nonumber\nonumber
\end{eqnarray}
\begin{eqnarray}
\,\,\,\,\,\,\,\,\,\,\,\,\,\,\,\,\,\,\,\,\,\,\,\,\,\,\,\,\,\,\,+\frac{e^2E^2\langle \delta z^2\rangle}{m_{\mathrm{d}}^2\beta\gamma\omega^2_0}\label{eq61}
\end{eqnarray}
\begin{figure}[t]
\begin{center}
\includegraphics[width=9cm,height=6.75cm]{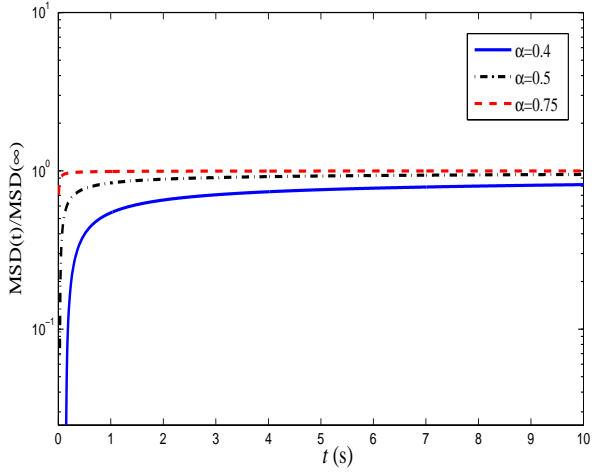}
\caption{\label{figure1} Normalized mean-square displacements of the dust as a function of time,  for $\gamma$=$1.54\times10^3\mathrm{s}^{-1}$, $\tau_{cor}$=$1.86\times10^{-4}\mathrm{s}$, $\omega_0$=$628\mathrm{s}^{-1}$, and $\alpha$=0.4, 0.5, 0.75 (solid line, dashed-dotted line, and dashed line, respectively).  }
\end{center}
\end{figure}
In figure~\ref{figure1}, we display the curves of the normalized MSD for the values $\gamma$=$1.54\times10^3\mathrm{s}^{-1}$, $\tau_{cor}$=$1.86\times10^{-4}\mathrm{s}$, $\omega_0$=$628\mathrm{s}^{-1}$, and $\alpha$=0.4, 0.5, 0.75. As we expected, in the limit of $t\to\infty$, the MSD approaches to the constant value
\begin{equation}
MSD(\infty)\sim‎‎‎‎‎‎‎‎‎‎\frac{e^2 E‎^2‎\langle‎‎\delta Z^2‎\rangle‎}{m^2_{\mathrm{d}}\beta‎\gamma‎‎\omega‎^2_0},\label{eq62}
\end{equation}
the constant value for the MSD is due to trapping of dust particle in a harmonic potential well; therefore, the diffusion of the dust particle occurs in a limit area. It is important to note here that the diffusion of a dust particle means a process of random displacements of a dust particle in a specified time interval due to charge fluctuations.
The equation~(\ref{eq62}) coincides with the result obtained from Markov dynamics because in the long times limit, the observation time $t$ is much longer than other characteristics time scales including the correlation time of the random force due to dust charge fluctuations $\tau_{cor}$ and the relaxation time of the dust velocity $\tau_{rel}$; as a result, there are no longer the memory effects in the velocity of the dust particle, and the dynamics of the dust particle in the long times limit is a Markov dynamics.

To calculate the dust temperature, by substituting $t_1$=$t_2$\\=$t$ into equation~(\ref{eq49}), and using ‎$‎\langle‎‎ v(t)\rangle$‎‎=$v_0g(t)-‎\omega‎^2_0x_0G(t)$, we obtain
\begin{eqnarray}
\langle v^2(t)\rangle=\left(v_0g(t)-‎\omega‎^2_0x_0G(t)‎‎\right)‎^2+‎\nonumber
\end{eqnarray}
\begin{equation}
\,\,\,\,\,\,\,\,\,\,\,\,\,\,\,\,\,\frac{e^2E^2‎\langle‎\delta ‎Z^2‎‎‎\rangle‎}{2m^2_{\mathrm{d}}\beta‎\gamma‎}‎‎\left(1-‎\omega‎^2_0G^2(t)-g^2(t)‎‎\right)‎\label{eq63}
\end{equation}
In the long-time limit, by substituting equations~(\ref{eq42}) and~(\ref{eq51}) into~(\ref{eq63}), we obtain the temperature of the dust particle as follows
\begin{equation}
T_{\mathrm{d}}=‎m_{\mathrm{d}}‎\langle v^2(t)‎‎\rangle‎\sim‎‎\frac{e^2E^2‎\langle‎\delta ‎Z^2‎‎‎\rangle‎}{2m_{\mathrm{d}}\beta‎\gamma‎},\label{eq64}
\end{equation}
which is consistent with the temperature obtained from reference~\cite{Quinn2000}, for the Markov velocity process .

\begin{figure}
\begin{center}
\includegraphics[width=9cm,height=6.75cm]{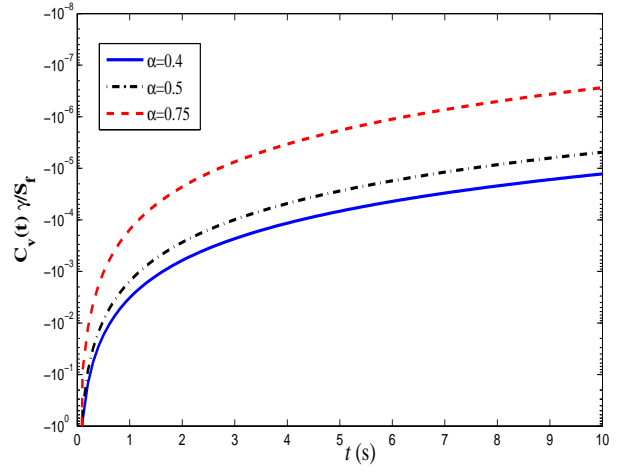}
\caption{\label{figure2} Normalized velocity autocorrelation functions of the dust as a function of time, for $\gamma$=$1.54\times10^3\mathrm{s}^{-1}$, $\tau_{cor}$=$1.86\times10^{-4}\mathrm{s}$, $\omega_0$=$62.8\mathrm{s}^{-1}$, and $\alpha$=0.4, 0.5, 0.75 (solid line, dashed-dotted line, and dashed line, respectively).  }
\end{center}
\end{figure}

Moreover, using equations~(\ref{eq58}),~(\ref{eq47})~(\ref{eq46}),~(\ref{eq59}), and some calculations, the long-time behavior of the velocity autocorrelation function of the dust particle can be written as
\begin{equation}
C_{\mathrm{v}}(t)\sim-‎\frac{e^2E^2‎\langle‎\delta ‎Z^2‎‎‎\rangle‎}{2m^2_{\mathrm{d}}\beta‎\gamma\omega^4_0‎}‎‎‎‎\frac{\gamma‎\tau_{cor}^{‎-3‎}}{‎\Gamma‎(-1-‎\alpha‎)}‎‎\left(‎\frac{t}{\tau_{cor}}‎‎‎\right)‎^{-\alpha‎-2}.\label{eq65}
\end{equation}
In contrast to the velocity autocorrelation function for the Markov velocity process of the dust particle obtained by equation~(\ref{eq23}), which decays oscillatory with time,
we observe from equation~(\ref{eq65}) that memory effects in the velocity process of the dust particle cause a power-law decay like $t^{-\alpha-2}$ for the velocity autocorrelation function of the dust particle, which indicates anomalous diffusion of the dust particle in dusty plasma. The power-law decay can obviously be seen in figure~\ref{figure2} that we show the curves $C_{\mathrm{v}}(t)$ for the values $\gamma$=$1.54\times10^3\mathrm{s}^{-1}$, $\tau_{cor}$=$1.86\times10^{-4}\mathrm{s}$, $\omega_0$=$62.8\mathrm{s}^{-1}$, and $\alpha$=0.4, 0.5, 0.75. For the given values of the parameter $\alpha$ , the function $\Gamma(-\alpha-1)$ is positive, and the curves $C_{\mathrm{v}}(t)$ have a negative tail at all times, $C_{\mathrm{v}}(t)<0$. The function $C_{\mathrm{v}}(t)$ is the average of the velocity of a dust particle at the time $t$ multiplied by its velocity at a later time. When $C_{\mathrm{v}}(t)$ is negative, this means that the dust particle diffuses in the opposite direction compared to $t$=0. As we expected, the function $C_v(t)$ of the dust particle decays to zero for the long times which is consistent with the result obtained from the Markov velocity process, because in long times $(t\to\infty)$, i.e., when observation time t is much longer than $\tau_{cor}$ and $\tau_{rel}$, there are no longer the memory effects in the velocity of the dust particle, and the dynamics of the dust particle in the long times limit is a Markov dynamics.

In conclusion, it is worth mentioning that dust particles collide with one another and depending on the grain relative velocity, coagulation or shattering processes can occur. In fact, there is a threshold velocity, $v_{\rm{shat}}$. If the grain relative velocity is smaller than $v_{\rm{shat}}$, coagulation occurs, i.e. grains collide and stick together. However, if the grain relative velocity is larger than $v_{\rm{shat}}$, shattering occurs, i.e. grains collide and shatter into smaller fragments~\cite{Hoang}. This is especially important in space dusty plasmas and astrophysical environments. Coagulation or shattering processes can affect the grain size distribution which, in turn, can affect the correlation time of dust charge fluctuations. Hence, these processes should be included in more realistic models.\\ 
\section{\label{summary}Summary and conclusions}
In this paper, by comparing the relevant timescales of a dusty plasma system, we have studied memory effects in the velocity process of an isolated dust particle exposed to the random force due to dust charge fluctuations in the sheath.  
We have presented a model based on the fractional Langevin equation to investigate the evolution of the dust particle and obtained a fluctuation-dissipation theorem for the dust grain.
Then, we have calculated the velocity, position, mean-square displacement, and the velocity autocorrelation function of the dust grain in terms of the generalized Mittag-Leffler functions and investigated the asymptotic behaviors of the MSD, VACF, and the dust temperature due to charge fluctuations in the long-time limit. We have found that in the presence of memory effects, the dust particle has the anomalous diffusion.
In this case, the velocity autocorrelation function of the dust grain decays as a power-law instead of the oscillatory (nonmonotonic) decay.\\
In this research, we have studied anomalous diffusion in laboratory dusty plasmas. Anomalous diffusion also occurs in astrophysical plasmas. An application of the mechanism of anomalous diffusion is to the formation of planets~\cite{Bracco1999,Barge1995,Tanga1996}. 
Also, anomalous diffusion can be found in the turbulent magnetotail~\cite{Zimbardo2010,Zimbardo2000,Chiaravalloti2006}, and solar wind turbulence~\cite{Zimbardo2005}, and heliosphere~\cite{Zimbardo2012}

\nocite{*}

\bibliography{Ghannad}

\end{document}